# From Trait to Behavior: A Cognitive-Affective Personality System (CAPS) Perspective on Multi-Homing Intention in AIGC Platforms

Xuchao Zhang, Jihye Lee

*School of Design, Pusan National University, Busan, South Korea*

**Abstract:** With the rapid development of Artificial Intelligence Generated Content (AIGC) platforms, users increasingly show cross-platform usage intentions. Existing research focuses on adoption and usage intentions in single-platform AIGC contexts. A theoretical gap still exists in studies on cross-platform usage. This paper constructs and verifies a three-stage multiple mediation model based on the personality trait–perception–behavioral response framework. The model integrates the optimum stimulation level (OSL) theory, complementarity theory, and perceived value theory, and it sets social influence and use experience as control variables to examine users' multi-homing intention. The results show that: (a) OSL significantly enhances users' perceived complementarity; (b) perceived complementarity positively affects perceived epistemic value; (c) perceived epistemic value significantly and positively predicts multi-homing intention; (d) OSL influences multi-homing intention through a chain mediation path of perceived complementarity and perceived epistemic value; and (e) social influence has a significant positive effect on multi-homing intention, while the effect of use experience is not significant.

## 1. Introduction

Artificial intelligence-generated content (AIGC) is the automated creation of text, images, videos, 3D assets, and other media using advanced AI algorithms, streamlining content production (Foo et al., 2025). Since the launch of representative AIGC platforms such as ChatGPT and DALL·E 2, the technology has seen widespread adoption across diverse fields, including art (Anantrasirichai & Bull, 2022), advertising (Kietzmann et al., 2018), and education (Kandlhofer et al., 2016). Recent statistics indicate that by early 2025, generative AI tools had 115–180 million daily active users worldwide. This figure is projected to grow by over 826 million by 2031, reaching about 1.2 billion users (Pohrebniyak, 2024). As AIGC platforms become increasingly differentiated in their functions, users are more likely to adopt multiple competing services to meet diverse content creation needs. This simultaneous use of various platforms, known as multi-homing (Mital & Sarkar, 2011), has become a central topic in research on cross-platform usage intentions (Gu et al., 2016; Manchanda & Deb, 2022; Sun et al., 2018).

Existing research focuses on users' migration intentions from traditional tools to AIGC (Liu & Ji, 2025), switching intentions between AIGC platforms (Zhou & Li, 2025), and usage intentions of AIGC tools (Lai et al., 2025; Ma & Huo, 2023; Skjuve et al., 2024; Zhou & Lu, 2025). However, these studies mostly start from single-platform adoption behavior, which makes it challenging to explain users' multi-homing intention for using multiple AIGC tools simultaneously. Some scholars have adopted the complementarity theory to describe users' multi-homing intention. Still, related research mainly focuses on SNS (Gu et al., 2016), B2B platforms (Manchanda & Deb, 2022), and accommodation platforms (Sun et al., 2018), with a lack of systematic theoretical construction and empirical testing for AIGC platforms.

To fill this research gap, this study constructs and validates a structural equation model focusing on AIGC multi-homing intention based on the personality trait-perception-behavioral response three-stage structure, integrating Optimum Stimulation Level (OSL) theory, complementarity theory, and perceived value theory to reveal the driving mechanisms of users' proactive formation of cross-platform usage intentions. Specifically, this study regards OSL as an antecedent variable of users' traits, explores how it drives users' sensitivity to functional complementarity among AIGC platforms, and examines how this complementarity perception indirectly affects multi-homing intention through perceived epistemic value.

## 2. Literature Review

### *2.1. Optimum Stimulation Level*

As Raju (1980), OSL is an individual's general tendency to respond to stimulus levels in specific situations. It is a psychological trait variable that describes how an individual requires external novelty, complexity, and variety in their environment. High OSL individuals are more stimulus-seeking and prefer novel, changing, or complex situations and activities (Kish & Donnenwerth, 1972). Conversely, individuals with low OSL tend to prefer familiar environments and are therefore more likely to exhibit withdrawal behavior in unfamiliar environments (Raju, 1980). Numerous studies have confirmed the critical role of OSL in behavioral selection. Individuals with higher OSL levels have a lower tolerance for repetitive stimuli. These individuals also tend to experience boredom in stable situations (Kish & Donnenwerth, 1972). Therefore, they seek out original, inconsistent, or intricate surroundings and actively explore behaviors (Vries et al., 2009). For example, exploring new tools or integrating cross-platform use to maintain optimal stimulation levels.

OSL is often used to analyze users' exploratory behavior tendencies and behavioral responses when faced with new technologies or products. It has been proven that OSL can predict exploring behavior towards new and innovative products (Steenkamp & Baumgartner, 1992), SNS multi-homing intention (Gu et al., 2016), and perceptions and usage intentions in the adoption of new technologies (Gu et al., 2018). Overall, these studies have demonstrated that OSL is a core psychological trait engendering users' adoption of new platforms and complex information environments.

Based on this, this study sets OSL as the antecedent variable in the model. It explores how it influences multi-homing intentions through a chain of mediating mechanisms involving users' perceived complementarity and perceived epistemic value among AIGC platforms.

### *2.2. Complementarity*

In this study, the concept of complementarity originates from Edgeworth and describes situations in which greater engagement in one activity increases the benefits of participating in related activities (Barua et al., 1996; Milgrom & Roberts, 1995). This relationship, in which the whole is greater than the sum of its parts, is known as complementarity theory. It has been applied to corporate strategic management (Barua et al., 1996), digital marketing (Koukova et al., 2008), e-commerce (Hwang & Oh, 2009), and mobile instant messaging (Arquero et al., 2017) to understand how to combine multiple resources to generate greater benefits.

In the context of digital platforms, perceived complementarity is defined as users' subjective perception of the synergistic advantages of two or more platforms. This combined use can bring benefits that exceed those offered by a single platform (Gu et al., 2016). In AIGC, users must complete various creative tasks, such as text and image generation, video editing, PowerPoint creation, design assistance, programming, and more. Since no single platform can meet all task requirements, users will find complementarity in the functions or services of different platforms (Sun et al., 2018). This complementarity improves both creative efficiency and the overall experience. It reflects the cognitive mechanism behind cross-platform usage intent.

To further explain this mechanism in the context of AIGC multi-homing intention, this study combines complementarity theory with task–technology fit, cognitive trust, and hedonic motivation theory, proposing three types of complementarity variables: complementarity in AIGC task–technology fit, complementarity in AIGC cognitive trust, and complementarity in AIGC hedonic motivation, which describe users' different needs for task fit, cognitive assurance, and hedonic stimulation in multi-platform usage.

#### *2.2.1. Complementarity in AIGC Technology–Task Fit*

Goodhue and Thompson (1995) proposed the concept of technology-task fit (TTF), which refers to the tendency of users to adopt a technology when its characteristics are highly compatible with their task characteristics, thereby achieving higher utilization efficiency and satisfaction. Since then, researchers have applied the TTF model to predict technology adoption and usage behavior in contexts such as mobile information systems (Gebauer et al., 2010), learning management platforms (D'Ambra et al., 2013; McGill & Klobas, 2009), and e-commerce environments (Klopping & McKinney, 2004).

In the context of AIGC, TTF denotes the user's perception of the platform's functionality in terms of its alignment with their task objectives. The diversity of tasks is met by utilizing multiple platforms. For example, a marketing strategy design that uses ChatGPT to generate descriptive prompts and combines them with DALL·E for visualization (Ivanov & Almeida, 2024; Wahid et al., 2023). This cross-platform usage indicates that users perceive that different platforms can complement each other regarding task support. This finding is consistent with the view put forward by TTF, namely that the complexity and diversity of tasks require technical capabilities to be aligned with them (Goodhue & Thompson, 1995). In light of these findings, this study delineates the complementarity in AIGC technology–task fit as the utilization of disparate AIGC functions by users to address diverse task requirements efficaciously.

#### *2.2.2. Complementarity in AIGC Cognitive Trust*

Trust is a psychological state that influences behavior, formed through an individual's past experiences and decisions (Rousseau et al., 1998). Trust research encompasses organizational behavior (Dirks & Ferrin, 2001), e-commerce and online transactions (Gefen et al., 2003), and the adoption and continued use of information systems (McKnight et al., 2002; Pavlou, 2003). Komiak and Benbasat (2006) distinguished trust's cognitive and affective dimensions in information technology. In the context of AIGC, Zhou and Lu (2025) further propose that cognitive trust refers to users' perception of AIGC's ability to provide reliable and accurate information. Affective trust relates to users' belief that the platform cares about their interests. This study evaluates the platform’s rational capabilities and emphasizes only cognitive trust.

Due to the inevitable occurrence of AI hallucinations in the AIGC (Ji et al., 2023), these hallucinations generate content that appears reasonable but is misleading, causing users to doubt the reliability of content generated by a single platform. Therefore, uncertainty about content will prompt users to adopt different platforms to verify its reliability. Based on this, this study proposes complementarity in AIGC cognitive trust, meaning that users can perceive content as more reliable and accurate by cross-validating outputs from multiple AIGC systems.

#### 2.2.3. Complementarity in AIGC Hedonic Motivation

Hedonic motivation refers to the pleasure and satisfaction users experience when using technology (Brown & Venkatesh, 2005). It predicts whether users will adopt or continue using a particular technology. In the context of IS, hedonic motivation is an essential predictor of user engagement and behavioral intent, especially when technology satisfies instrumental needs and triggers pleasure and curiosity (Thong et al., 2006). Unlike extrinsic motivations, which focus on efficiency, hedonic motivations focus on psychological satisfaction and pleasure.

Users typically expect AIGC platforms to provide inspiration or novel experiences for creative generation tasks (Skjuve et al., 2024). Differences in output formats and interaction modes can influence this pleasurable experience. Therefore, combining multiple platforms can enhance overall pleasure and creative satisfaction compared to a single platform. Based on this, this study proposes complementarity in AIGC hedonic motivation, namely that users believe that a combination of multiple platforms can provide richer inspiration, expression, and immersion.

According to the Theory of Reasoned Action (TRA), behavior is driven by behavioral intention, and intention originates from attitude and subjective norms. In the TPB extension, Ajzen (1991) pointed out that stable individual difference variables can serve as background variables that indirectly influence behavioral intention by affecting the belief formation process. Recent studies have shown that OSL increases users’ sensitivity to external information, technology, and resources, and in multi-platform environments, enhances the motivation to identify and integrate complementary features (Gu et al., 2016). In AIGC usage, high OSL users are likelier to perceive differences among platforms in task support, output trust, and emotional experience, forming multi-platform complementarity cognition. Based on this, the following hypotheses are proposed:

**H1**: OSL positively influences perceived complementarity in AIGC technology–task fit.
**H2**: OSL positively influences perceived complementarity in AIGC cognitive trust.
**H3**: OSL positively influences perceived complementarity in AIGC hedonic motivation.

### 2.3. Perceived Epistemic Value

Perceived value is consumers' overall evaluation of a product based on comparing its benefits and costs (Zeithaml, 1988). Sheth et al. (1991) classified it into functional, conditional, social, emotional, and epistemic value. Some scholars also describe consumer perceived value as perceived utility value, perceived hedonic value, perceived social value, and perceived cognitive value (Sweeney & Soutar, 2001).

Epistemic value means the novelty a consumer perceives in a new product (Pihlström & Brush, 2008). In a mobile service context, it also includes curiosity about new content and the knowledge gained from trying new services (Pihlström & Brush, 2008). Therefore, in the AIGC context, perceived epistemic value refers to the novelty, curiosity, and knowledge users gain when engaging with generative content services. This study focuses on this dimension because it is a core cognitive mechanism driving exploration and adoption behavior in multi-platform usage contexts. Previous research supports that novelty is a key antecedent variable influencing users' adoption of AIGC technology (Ma & Huo, 2023). In addition, Skjuve et al. (2024), based on U&G theory, found that about 50% of users developed exploratory motivation due to the novelty of AIGC, and most participants indicated that their initial use of AIGC was driven by curiosity. Learning and development were also important drivers. These results confirm the key role of perceived epistemic value in AIGC usage intentions. Therefore, this study regards it as a core variable at the perception level and reveals the cognitive mechanisms behind multi-platform usage intentions driven by novelty, curiosity, and knowledge gain.

When the combination of AIGC functions matches the task, it can increase perceived value and promote epistemic exploration. Consistent with TTF logic, when technology effectively supports tasks, it can enhance the perception of performance impacts, namely system efficiency, quality, and satisfaction (Goodhue & Thompson, 1995). Previous studies have confirmed that technology-task fit significantly impacts perceived value (Zhang et al., 2021).

**H4**: Complementarity in AIGC Technology–Task Fit positively influences perceived epistemic value.

Users' combined trust in AIGC platforms can also significantly enhance their perceived value. Konuk (2018) found that brand trust significantly improved consumers' perceived value in retail. In AIGC, users may form differentiated cognitive trust based on the performance of different platforms. By combining multiple platforms, mutual verification can produce a complementary trust effect, enhancing users' perception of epistemic value.

**H5**: Complementarity in AIGC Cognitive Trust positively influences perceived epistemic value.

Wu et al. (2025), based on UTAUT2, perceived value theory, and SOR theory, found that hedonic motivation significantly enhances perceived value. In AIGC, different platforms can stimulate emotional pleasure through interaction methods, content styles, or creative feedback mechanisms. This hedonic motivation complementarity formed by such differences may further promote perceived epistemic value.

**H6**: Complementarity in AIGC Hedonic Motivation positively influences perceived epistemic value.

In summary, perceived epistemic value has been confirmed as a key antecedent variable driving users' adoption and continued use of AIGC technology (Ma & Huo, 2023; Skjuve et al., 2024). However, existing research mainly focuses on single-platform adoption behavior, with insufficient exploration of its role in multi-platform contexts. This study proposes that users can obtain stronger cognitive stimulation through combined usage in a multi-platform environment with significant differences, significantly enhancing perceived epistemic value. Perceived epistemic value influences users' adoption decisions for a single platform and plays a key role in multi-platform usage decisions. Therefore, this study proposes the following hypothesis and constructs the theoretical research structural model (Figure 1).

**H7**: Perceived epistemic value positively influences multi-homing intention.

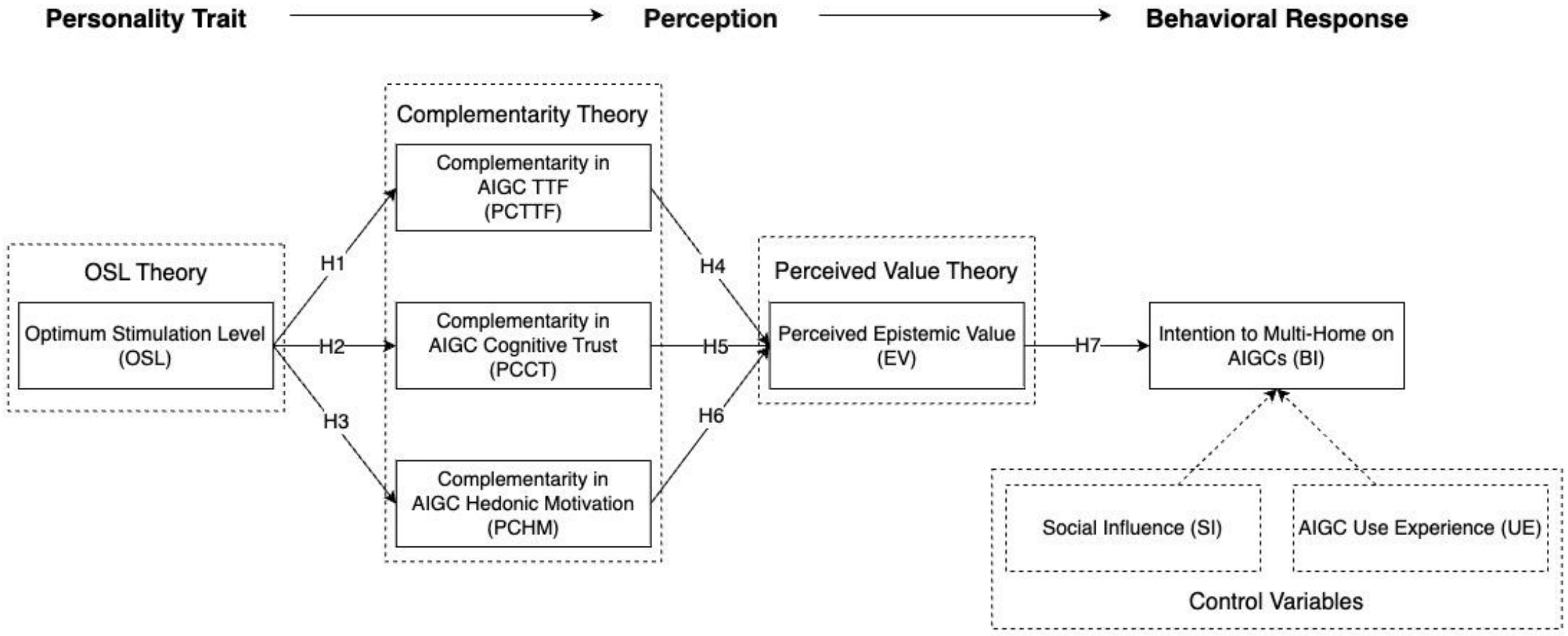

Figure 1. The research model

Table 1. Demographic information of the participants

| | Category | Number | Percentage |
|---|---|---|---|
| Gender | Male | 180 | 45.8 |
| | Female | 213 | 54.2 |
| Age | 18–25 | 160 | 40.7 |
| | 26–30 | 141 | 35.9 |
| | 31–40 | 69 | 17.6 |
| | 41–50 | 18 | 4.6 |
| | ⩾51 | 5 | 1.3 |
| Education | High school / Vocational school or below | 27 | 6.9 |
| | Associate degree | 37 | 9.4 |
| | Undergraduate | 180 | 45.8 |
| | Master | 88 | 22.4 |
| | PhD | 61 | 15.5 |
| Profession | Student | 216 | 55.0 |
| | Teacher / Researcher | 41 | 10.4 |
| | Creative Professional | 33 | 8.4 |
| | Corporate Employee | 52 | 13.2 |
| | Other | 51 | 13.1 |
| Number of AIGCs regularly used | 1 | 61 | 15.5 |
| | 2–3 | 167 | 42.5 |
| | 4–5 | 64 | 16.3 |
| | ⩾6 | 101 | 25.7 |
| Total | | 393 | 100 |

## 3. Method

### *3.1. Participants*

This study adopted a questionnaire survey to validate the proposed research model. We anonymously distributed and collected 618 questionnaires from Chinese-speaking users in China and Korea, with all participants participating voluntarily. Respondents received a cash incentive of 3 RMB to encourage questionnaire completion.

After excluding respondents who had not used AIGC platforms and those who failed the attention check questions, 393 valid questionnaires were retained. Among the respondents, 45.8% were male and 54.2% were female. A total of 76.6% were aged between 18 and 30 years. Moreover, 83.7% had obtained or were pursuing an undergraduate degree or higher education, and 86.9% belonged to the higher education or white-collar workforce group. Consistent with this, Lane (2024), in its report, noted that the core audience for AIGC is concentrated in the higher education demographic and information-intensive white-collar occupations, with educational attainment identified as one of the key factors determining exposure to AIGC technologies. Therefore, the sample characteristics in this study are highly consistent with the user profile described by an internationally recognized authority, providing strong representativeness and explanatory power (Table 1).

### *3.2. Instruments Development*

The questionnaire used in this study consisted of two sections. The first section collected participants' demographic information. The second section contained six core constructs and two control variables, totaling 26 items. To ensure content validity, all measurement items were adapted from well-established scales in prior literature and were uniformly measured using five-point Likert scales.

OSL items were adapted from Steenkamp and Baumgartner (1995). Perceived complementarity descriptions were adapted from Gu et al. (2016) and further refined into three dimensions: complementarity in AIGC task–technology fit, adapted from Howard and Rose (2019); complementarity in AIGC cognitive trust, adapted from Choi and Ji (2015); and complementarity in AIGC hedonic motivation based on the UTAUT2 model proposed by Venkatesh et al. (2012). Perceived epistemic value items were adapted from Pihlström and Brush (2008). Intention to multi-home on AIGCs items was adapted from Gu et al. (2016). In addition, this study included two control variables: social influence and AIGC use experience, adapted from Taylor and Todd (1995) and Dishaw and Strong (1999), respectively. To improve the readability and content validity of the questionnaire, a preliminary test was conducted with 50 users after the questionnaire design was completed, and language adjustments were made based on feedback to eliminate potential ambiguities and misunderstandings.

## 4. Results

### *4.1. The Measurement Model*

The measurement model's reliability and validity were evaluated through standardised factor loadings, Cronbach's alpha coefficients, convergent validity, discriminant validity, and model fit.

According to Schumacker and Lomax (2008), the standardised factor loadings of items should be at least 0.50. In this study, all items' factor loadings met this standard. To further improve the structural fit, one item from complementarity in AIGC technology–task fit was removed (Segars, 1997). Cronbach's alpha coefficients of the six core constructs and two control variables were all above 0.80 (0.800 to 0.840) (Table 2), indicating an excellent internal consistency.

Convergent validity was assessed using composite reliability (CR) and average variance extracted (AVE). Fornell and Larcker (1981) state that CR should be no lower than 0.70 for each construct, and AVE should be above 0.50. All constructs in this study meet these requirements (Table 2). For discriminant validity, the square root of AVE should be greater than the correlation coefficients between the construct and any other construct (Chin, 1998). Although some correlations were near the Fornell–Larcker threshold, the model showed adequate discriminant validity. The constructs came from different theories, which reduces concerns about conceptual overlap (Table 3).

Following the criteria proposed by Hu and Bentler (1999), the results show that the measurement model achieves satisfactory fit with $\chi^2/df$ (chi-square/degree of freedom) = 1.824, GFI (goodness-of-fit index) = 0.917, TLI (Tucker-Lewis index) = 0.960, CFI (comparative fit index) = 0.967, RMR (root mean square residual) = 0.033, RMSEA (root mean square error of approximation) = 0.046 (Table 4).

Table 2. Results of construct validity and reliability analysis

| Latent Variable | Measurement Variable | Mean | Std. Dev | Factor Loadings | α | CR | AVE |
|---|---|---|---|---|---|---|---|
| OSL | OSL1 | 3.90 | .86 | .744 | .835 | .839 | .567 |
| | OSL2 | | | .725 | | | |
| | OSL3 | | | .831 | | | |
| | OSL4 | | | .707 | | | |
| PCTTF | PCTTF1 | 4.07 | .85 | .732 | .813 | .813 | .593 |
| | PCTTF2 | | | .779 | | | |
| | PCTTF3 | | | .797 | | | |
| PCCT | PCCT1 | 4.06 | .83 | .727 | .807 | .809 | .585 |
| | PCCT2 | | | .807 | | | |
| | PCCT3 | | | .759 | | | |
| PCHM | PCHM1 | 3.97 | .86 | .762 | .839 | .841 | .638 |
| | PCHM2 | | | .822 | | | |
| | PCHM3 | | | .811 | | | |
| EV | EV1 | 4.06 | .86 | .806 | .817 | .818 | .600 |
| | EV2 | | | .778 | | | |
| | EV3 | | | .738 | | | |
| BI | BI1 | 4.04 | .92 | .854 | .800 | .820 | .612 |
| | BI2 | | | .561 | | | |
| | BI3 | | | .890 | | | |
| SI | SI1 | 3.89 | .90 | .814 | .802 | .839 | .634 |
| | SI2 | | | .793 | | | |
| | SI3 | | | .782 | | | |
| UE | UE1 | 3.81 | .91 | .762 | .840 | .801 | .574 |
| | UE2 | | | .817 | | | |
| | UE3 | | | .689 | | | |

*OSL* Optimum Stimulation Level, *PCTTF* Complementarity in AIGC Technology - Task Fit, *PCCT* Complementarity in AIGC Cognitive Trust, *PCHM* Complementarity in AIGC Hedonic Motivation, *EV* Perceived Epistemic Value, *BI* Multi-Homing Intention, *SI* Social Influence, *UE* AIGC Use Experience

Table 3. Discriminant validity of the research model

| Constructs | OSL | PCTTF | PCCT | PCHM | EV | BI | SI | UE |
|---|---|---|---|---|---|---|---|---|
| OSL | **.753** | | | | | | | |
| PCTTF | .612 | **.770** | | | | | | |
| PCCT | .558 | .785 | **.765** | | | | | |
| PCHM | .586 | .663 | .753 | **.799** | | | | |
| EV | .625 | .763 | .767 | .744 | **.775** | | | |
| BI | .544 | .634 | .688 | .644 | .693 | **.782** | | |
| SI | .552 | .595 | .688 | .658 | .697 | .645 | **.796** | |
| UE | .525 | .572 | .621 | .578 | .588 | .571 | .663 | **.758** |

Table 4. The goodness of fit indices for the measurement model and research model

| Model | $X^2$ | X2/df | GFI | TLI | CFI | RMR | RMSEA |
|---|---|---|---|---|---|---|---|
| Measurement model | 450.55 | 1.824 | .917 | .960 | .967 | .033 | .046 |
| Research model | 716.926 | 2.726 | .863 | .917 | .927 | .047 | .066 |
| Recommended criteria | $p > .05$ | < 5.0 | > .80 | > .90 | > .90 | < .05 | < .08 |

Table 5. The results of the hypothesis test

| Hypotheses | Hypothesized path | B | $\beta$ | S. E | $t$ | Result |
|---|---|---|---|---|---|---|
| H1 | OSL → PCTTF | 1.036 | .924 | .084 | 12.267*** | Supported |
| H2 | OSL → PCCT | 1.041 | .959 | .084 | 12.437*** | Supported |
| H3 | OSL → PCHM | 1.050 | .892 | .086 | 12.184*** | Supported |
| H4 | PCTTF → EV | .459 | .420 | .121 | 3.795*** | Supported |
| H5 | PCCT → EV | .338 | .299 | .138 | 2.445* | Supported |
| H6 | PCHM → EV | .316 | .304 | .088 | 3.608*** | Supported |
| H7 | EV → BI | .613 | .577 | .079 | 7.742*** | Supported |
| Control Variable | SI → BI | .266 | .261 | .091 | 2.915** | Supported |
| Control Variable | UE → BI | .108 | .106 | .082 | 1.321 | Rejected |

*OSL* Optimum Stimulation Level, *PCTTF* Complementarity in AIGC Technology–Task Fit, *PCCT* Complementarity in AIGC Cognitive Trust, *PCHM* Complementarity in AIGC Hedonic Motivation, *EV* Perceived Epistemic Value, *BI* Multi-Homing Intention, *SI* Social Influence, ***$p$<.001, **$p$<.01, *$p$<.05

## 4.2. The Structural Model

SEM was applied to test the research model. The model showed acceptable fit with $\chi^2$/df = 2.726, GFI = 0.863, TLI = 0.917, CFI = 0.927, RMR = 0.047, RMSEA = 0.066 (Table 4). The testing results supported all seven hypotheses (Table 5).

OSL is positively associated with complementarity in AIGC task–technology fit ($\beta$ = 0.924, $p$ = 0.000), and complementarity in AIGC cognitive trust ($\beta$ = 0.959, $p$ = 0.000) and complementarity in AIGC hedonic motivation ($\beta$ = 0.892, $p$ = 0.000), supporting hypotheses H1, H2, and H3. Complementarity in AIGC task–technology fit is positively associated with perceived epistemic value ($\beta$ = 0.420, $p$ = 0.000), as are complementarity in AIGC cognitive trust ($\beta$ = 0.299, $p$ = 0.014) and complementarity in AIGC hedonic motivation ($\beta$ = 0.304, $p$ = 0.000), supporting hypotheses H4, H5, and H6. Perceived epistemic value is significantly associated with multi-homing intention ($\beta$ = 0.577, $p$ = 0.000), supporting H7. In addition, among the control variables, social influence positively affects multi-homing intention ($\beta$ = 0.261, $p$ = 0.004), whereas the effect of use experience is insignificant ($\beta$ = 0.106, $p$ = 0.186). The verified research structural model is shown in Figure 2.

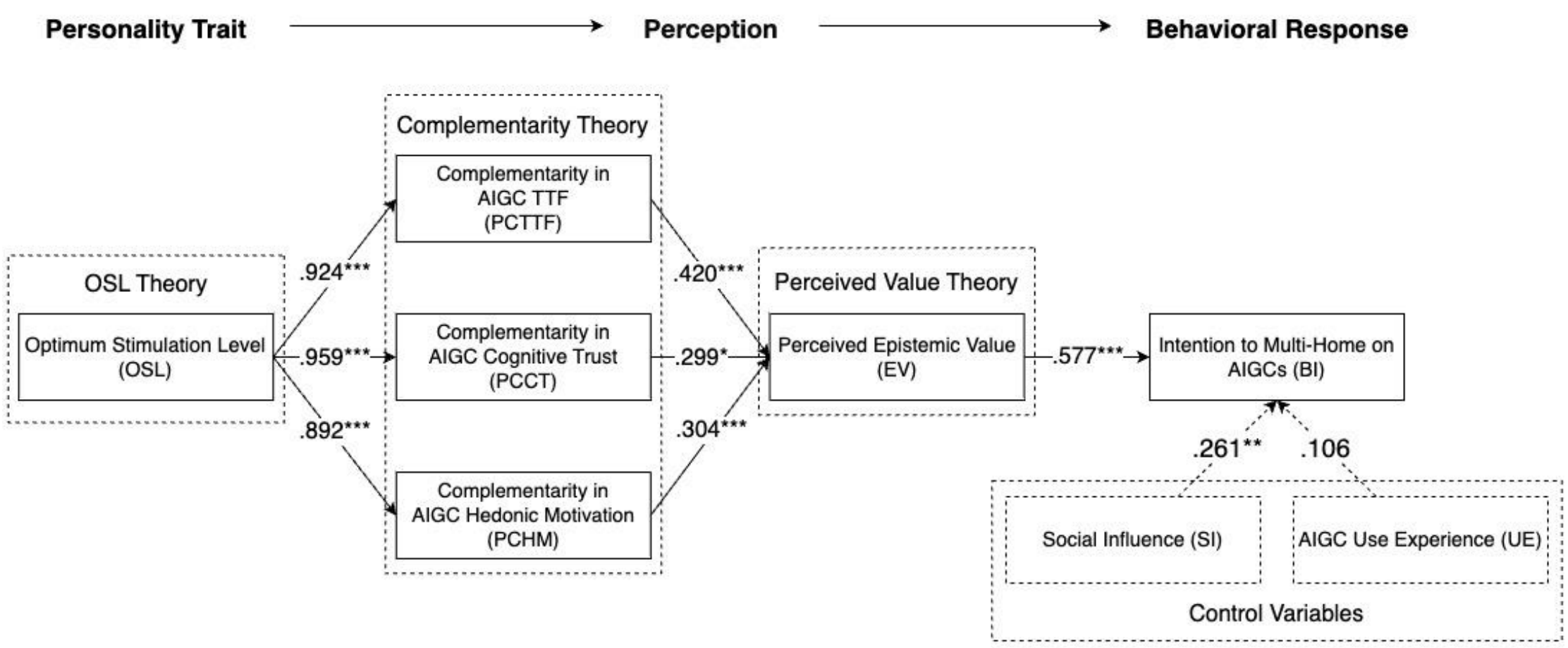

Figure 2. The research model with its standardised coefficients. ***$p$ < .001, **$p$ < .01, *$p$ < .05

Table 6. Direct, indirect, and total effects among the variables

| Dependent variable | Independent variable | Direct effect | Indirect effect | Total effect |
| --- | --- | --- | --- | --- |
| PCTTF | OSL | 0.924*** | _ | .924 |
| PCCT | OSL | 0.959*** | _ | .959 |
| PCHM | OSL | 0.892*** | _ | .892 |
| EV | OSL | _ | .946 | .946 |
| | PCTTF | 0.42*** | _ | .420 |
| | PCCT | 0.299* | _ | .299 |
| | PCHM | 0.304*** | _ | .304 |
| BI | OSL | _ | .546 | .546 |
| | PCTTF | _ | .242 | .242 |
| | PCHM | _ | .175 | .175 |
| | PCCT | _ | .172 | .172 |
| | EV | 0.577*** | _ | .577 |
| | SI | 0.261** | _ | .261 |
| | UE | .106 | _ | .106 |

*OSL* Optimum Stimulation Level, *PCTTF* Complementarity in AIGC Technology–Task Fit, *PCCT* Complementarity in AIGC Cognitive Trust, *PCHM* Complementarity in AIGC Hedonic Motivation, *EV* Perceived Epistemic Value, *BI* Multi-Homing Intention, *SI* Social Influence, ***$p$ < .001, **$p$ < .01, *$p$ < .05

## *4.3. Direct, Indirect, and Total Effects*

Table 6 presents the direct, indirect, and total effects among the latent variables in the structural model.

First, the OSL significantly and directly influenced three types of perceived complementarity: complementarity in AIGC task–technology fit ($\beta$ = 0.924, $p$ = 0.000), complementarity in AIGC cognitive trust ($\beta$ = 0.959, $p$ = 0.000), and complementarity in AIGC hedonic motivation ($\beta$ = 0.892, $p$ = 0.000).

Second, these three types of perceived complementarity significantly predicted perceived epistemic value, complementarity in AIGC task–technology fit ($\beta$ = 0.420, $p$ = 0.000), complementarity in AIGC cognitive trust ($\beta$ = 0.299, $p$ = 0.014), and complementarity in AIGC hedonic motivation ($\beta$ = 0.304, $p$ = 0.000). Notably, OSL also showed a strong indirect effect on perceived epistemic value (indirect effect = 0.946), implying a complete mediation via complementarity perceptions.

Third, multi-homing intention was significantly influenced by perceived epistemic value ($\beta$ = 0.577, $p$ = 0.000), as well as by two control variables: social influence ($\beta$ = 0.261, $p$ = 0.004) and use experience ($\beta$ = 0.106, $p > 0.05$). Among the indirect paths, OSL exhibited a substantial total indirect effect on multi-homing intention ($\beta$ = 0.546), which was primarily mediated through the pathway OSL → complementarity in AIGC task–technology fit / complementarity in AIGC cognitive trust / complementarity in AIGC hedonic motivation → perceived epistemic value → multi-homing intention.

## 5. Discussion

### *5.1. The Effects of OSL on Perceived Complementarity*

The results show that OSL has a significant positive effect on all three types of perceived complementarity, with the most potent predictive effect on complementarity in AIGC cognitive trust, followed by complementarity in AIGC task–technology fit and complementarity in AIGC hedonic motivation. This finding supports the core proposition of Stimulus-Seeking Theory that individuals with high OSL are more inclined to seek novel, complex, and diverse experiences (Steenkamp & Baumgartner, 1995) and are thus more sensitive to functional differences and complementarity among platforms. This result is consistent with Gu et al. (2016), who found that OSL significantly enhances users' sensitivity to platform complementarity features.

First, OSL significantly positively impacts the complementarity in AIGC task-technology fit, indicating that users with high OSL can better recognize the complementarity of technological capabilities across different platforms. This is consistent with TTF logic, whereby individuals seek out technologies to improve task efficiency when faced with diverse task requirements (Goodhue & Thompson, 1995). High OSL users need novel or complex environmental stimuli and feel limited by the functionality of a single platform. Therefore, they actively explore the advantages of different platforms to complete tasks efficiently.

Second, OSL has the most potent predictive effect on complementarity in AIGC cognitive trust, indicating that users with high OSL are particularly concerned about the accuracy and reliability of platform outputs and tend to enhance overall cognitive trust through a combination of multiple platforms. This is consistent with the findings of Gu et al. (2016), namely that individuals with high OSL can recognize environmental differences and integrate multiple sources of information. In the AIGC context, due to the risk of AI hallucinations, high OSL users leverage complementary trust across different platforms to enhance the accuracy and reliability of content.

Finally, the significant effect of OSL on complementarity in AIGC hedonic motivation shows that high OSL users not only focus on functionality and trust but also value emotional stimulation and creative pleasure. Based on UTAUT2, hedonic motivation is a key driver of technology adoption (Venkatesh et al., 2012). The interaction modes and output styles on different platforms provide high OSL users with complementary emotional experiences, encouraging them to achieve greater creative satisfaction through a combination of multiple platforms.

### *5.2. The Impact of Perceived Complementarity on Perceived Epistemic Value*

The results show that all three types of perceived complementarity significantly positively affect perceived epistemic value. Among them, complementarity in AIGC task–technology fit has the most potent predictive effect, followed by complementarity in AIGC hedonic motivation and complementarity in AIGC cognitive trust. This indicates that, in multi-platform usage, the complementary characteristics perceived by users in functional, trust, and emotional dimensions can effectively enhance their subjective evaluation of the novelty, curiosity, and knowledge gain of platform content. This finding is consistent with Gu et al. (2016), namely that platform complementarity can enhance users' perception of added value.

First, the most substantial effect of complementarity in the AIGC task–technology fit can be explained by TTF, which posits that a high task–technology match reduces cognitive load and improves task efficiency (Goodhue & Thompson, 1995). AIGC platforms offer different specialized services. Users can better unleash their creativity by using these specialized platforms. This is consistent with the findings of Zhang et al. (2021), namely that when the functionality of digital content is highly matched to the task, users' perceived value will be significantly enhanced.

Second, the positive effect of complementarity in AIGC cognitive trust indicates that the accuracy and reliability of platform output results are essential prerequisites for forming epistemic value. Under the risk of AI hallucinations, users enhance the consistency of information through cross-platform verification, thereby strengthening their motivation to acquire knowledge. Previous studies have pointed out that trust is a psychological prerequisite for continuous exploration and information acquisition in the digital content environment (Gefen et al., 2003; Konuk, 2018), and its importance is also highlighted in multi-platform usage scenarios.

Finally, the positive effect of complementarity in AIGC hedonic motivation shows the role of emotional experience in the formation of cognitive value. Wu et al. (2025) noted that hedonic motivation can significantly enhance users’ perceived value of platform content. This conclusion also applies to AIGC multi-platform usage scenarios. Differences between platforms in terms of interface design, interaction feedback, and content style can significantly enhance users' desire to explore and their sense of immersion.

### *5.3. The Role of Perceived Epistemic Value in Promoting Multi-Homing Intention*

The results show that perceived epistemic value has a significant positive effect on multi-homing intention. This is consistent with the findings of Ma and Huo (2023) and Skjuve et al. (2024). This shows that users' subjective evaluations of multi-platform novelty, curiosity, and knowledge gain are key drivers of their intention to use multi-platform services.

In a multi-platform context, perceived epistemic value significantly impacts multi-platform usage intention as an extension of perceived complementarity. Even when social influence and use experience are controlled for, the predictive role of perceived epistemic value remains significant, indicating its stability and independence in forming behavioral intentions. This finding suggests that, rather than relying solely on social dissemination or the accumulation of user experience, platform operators should stimulate cross-platform usage intention by leveraging the cognitive stimulation brought by multi-platform combinations.

In summary, perceived epistemic value directly predicts multi-homing intention and is the core cognitive mechanism linking complementarity perception to behavioral intention. Within the three-stage pathway structure proposed in this study, it fully embodies the logical chain of personality trait–perception–behavioral response.

### 5.4. Indirect Effects and Mediation Pathways

This study further examined the indirect effect pathways of OSL on multi-homing intention. The results indicate that OSL affects multi-homing intention entirely via indirect pathways formed through three types of perceived complementarity and the resulting perceived epistemic value. In this process, OSL first enhances perceptions of complementarity in multi-platform combinations, which then, through the cognitive transformation of perceived epistemic value, promote the formation of behavioral intention. This mechanism illustrates the complete route by which a personality trait is transmitted to a behavioral response via perception, highlighting the applicability of the personality trait–perception–behavioral response logic in the AIGC multi-platform usage context and supporting the theoretical alignment and empirical robustness of the three-stage multiple mediation model proposed in this study.

### 5.5. The Role of Control Variables

In this study, we included social influence and use experience as control variables to test the relationship of these two constructs with multi-homing intention. Findings of the analysis indicate that social influence positively affects multi-homing intention, but the impact of use experience is not essential. This result suggests that, in the AIGC multi-platform usage context, users' multi-homing intention is more likely to be driven by external social factors such as peers, industry trends, or social recommendations, rather than relying entirely on their existing use experience for judgment. This finding further emphasizes that external social norms may have stronger predictive power when platform differences are significant than individual experience. Especially in the rapid development of AIGC, users are more inclined to shape their platform preferences by referring to others' opinions and mainstream viewpoints. Therefore, platform operators should focus on shaping the social influence mechanism, such as strengthening word-of-mouth dissemination, expert endorsements, and community guidance strategies, to enhance users' platform usage intentions.

### 5.6. Implications for Practice

This study developed and empirically tested a model of users' multi-homing intention on AIGC platforms, providing three practical implications for platform operation and technology development. First, high OSL users are more sensitive to platform differences and combine multiple platforms to satisfy cognitive motivation. Platforms can identify such exploratory users through behavioral data and offer strategies such as cross-platform task distribution and joint subscription packages to enhance overall collaborative value. Second, perceived complementarity significantly enhances epistemic value and multi-homing intention, indicating that users prefer platform complementarity over reliance on a single platform. Platforms should reduce closed ecosystems and promote API integration, account interoperability, and collaborative recommendation mechanisms to form an open and symbiotic usage model. Third, perceived epistemic value is the core psychological mechanism of multi-homing intention. Platform content should improve creativity, unexpectedness, and diversity, while optimizing deep feedback mechanisms to enhance users' cognitive engagement and willingness for continued use.

## 6. Conclusion

This study constructed and validated a three-stage multiple mediation model to explain users' multi-homing intention on AIGC platforms. The study focused on how OSL indirectly influences multi-homing intention through users' perceptions of platform complementarity features and epistemic value. The model depicts complementarity perception from task-technology fit, cognitive trust, and hedonic motivation. It introduces perceived epistemic value as a mediating variable, revealing the multi-stage cognitive mechanism in users' combined platform usage.

The main theoretical contributions of this study are as follows:

First, it proposes and validates a three-stage multiple mediation pathway model based on personality trait–perception–behavioral response. This model reveals the complete pathway from OSL → perceived complementarity → perceived epistemic value → multi-homing intention, providing a quantifiable theoretical framework for understanding AIGC users' multi-platform adoption behavior.

Second, it extends the application boundaries of OSL theory, complementarity theory, and perceived value theory in AIGC research. Integrating TTF, cognitive trust, and hedonic motivation clarifies the structural dimensions of complementarity perception, providing a foundation for theorization and measurement in multi-platform contexts.

Third, it verifies the role of perceived complementarity and perceived epistemic value in chain mediation. The results show that OSL indirectly drives behavioral intention by stimulating complementarity perception in multi-platform combinations and transforming it into epistemic value. This finding provides a new theoretical anchor for integrating user psychological traits with research on multi-platform interaction behavior.

This study has three main limitations. First, it adopted a cross-sectional questionnaire design, which cannot capture the dynamic changes in users' cognitive mechanisms and platform usage behaviors. Multi-homing intention may be continuously influenced by changes over time and in platform functions, indicating a process-oriented nature. Future research could employ longitudinal tracking or experimental designs to verify the causal stability and stage-specific differences in the pathway relationships. Second, although this study differentiated tasks–technology fit, cognitive trust, and hedonic motivation in complementarity perception- it did not consider the moderating effects of individual characteristics such as technology anxiety, switching costs, or platform dependence. These variables may influence users' sensitivity to complementarity and the construction pathway of their perceived epistemic value. Future research could introduce individual-level variables as moderators. Third, this study focused on the general usage context of AIGC platforms without distinguishing specific platform types or task objectives. The complementary structures and cognitive logics may vary due to the significant differences among current AIGC platforms in content types and usage goals. Future research could develop sub-models tailored to specific platforms or usage scenarios for group testing, thereby enhancing the explanatory power and applicability of the model.

**Acknowledgement**
This study received no specific grant from any funding agency in the public, commercial, or not-for-profit sectors.

**Declaration of Interest statement**
The authors report there are no competing interests to declare.

**Data availability statement**
The data used in this study are available from the corresponding author upon reasonable request.

**Ethical approval**
This research was reviewed and approved by the Pusan National University Institutional Review Board (PNU IRB) and was classified as Exempt Review. The approval reference number is *PNU IRB 2025-09-003*. According to the exemption determination, written consent was waived because the study used a survey method that did not include personally identifiable information. Informed consent was obtained from all participants prior to data collection.

**CRediT authorship contribution statement**
Xuchao Zhang: Data curation, Formal analysis, Investigation, Methodology, Software, Validation, Visualization, Writing – original draft.
Jihye Lee: Conceptualization, Funding acquisition, Project administration, Resources, Supervision, Writing – review and editing.